\documentclass[11pt]{article}
\usepackage{times}
\usepackage{geometry}
\usepackage[utf8]{inputenc}
\usepackage{enumitem,amssymb}
\usepackage{graphicx}
\usepackage{fancyhdr}
\usepackage{aas_macros}
\usepackage{mdframed}
\usepackage{wrapfig}

\usepackage[authoryear]{natbib}
\setcitestyle{authoryear,open={(},close={)}}

\mdfdefinestyle{theoremstyle}{
innertopmargin=\topskip,}
\mdtheorem[style=theoremstyle]{lrptextbox}{}

\title{\huge{Revealing the Origin and Cosmic Evolution of Supermassive Black Holes}}
\author{T. E. Woods$^{1}$, R. M. Alexandroff$^{2}$, S. L. Ellison$^{3}$, L. Ferrarese$^{1}$, S. C. Gallagher$^{4}$, L. Gallo$^{5}$, D. Haggard$^{6}$,\\ P. B. Hall$^{7}$, J. Hlavacek-Larrondo$^{8}$, V. C. Khatu$^{4}$, A. W. S. Man$^{2}$, S. McGee$^{9}$, B. R. McNamara$^{10}$, J. Ruan$^{6}$,\\ G. Sivakoff$^{11}$, I. H. Stairs$^{12}$, C. Willott$^{1}$\\
{\small$^{1}$ National Research Council, Herzberg Astronomy \& Astrophysics, 5071 West Saanich Road, Victoria, BC V9E 2E7, Canada}\\
{\small$^{2}$Dunlap Institute for Astronomy and Astrophysics,University of Toronto, 50 St. George Street, Toronto, ON, M5S 3H4, Canada}\\
{\small$^{3}$Department of Physics and Astronomy, University of Victoria, Victoria, British Columbia, V8P 1A1, Canada}\\
{\small$^{4}$Department of Physics \& Astronomy, Western University, London, ON, N6A 3K7, Canada}\\
{\small$^{5}$Department of Astronomy \& Physics, Saint Mary’s University, 923 Robie Street, Halifax, Nova Scotia, B3H 3C3, Canada}\\
{\small$^{6}$McGill Space Institute and Department of Physics, McGill University, 3600 rue University, Montreal, QC, H3A 2T8, Canada}\\
{\small$^{7}$Department of Physics and Astronomy, York University, Toronto, ONM3J 1P3, Canada}\\
{\small$^{8}$D{\'e}partement de Physique, Universit{\'e} de Montr{\'e}al, Montr{\'e}al QC H3C 3J7, Canada}\\
{\small$^{9}$School of Physics and Astronomy, University of Birmingham, Edgbaston, Birmingham, B15 2TT, UK}\\
{\small$^{10}$Department of Physics and Astronomy, University of Waterloo, Waterloo, ON N2L 3G1, Canada}\\
{\small$^{11}$Department of Physics, CCIS 4-183, University of Alberta, Edmonton, AB, T6G 2E1, Canada}\\
{\small$^{12}$Dept. of Physics and Astronomy, University of British Columbia, 6224 Agricultural Road, Vancouver, B.C., V6T 1Z1, Canada}\\
}

\date{White Paper submitted to LRP2020 panel on 30th September 2019}
\begin{document}
\setlength{\bibsep}{0.0pt}
\maketitle


\noindent The next generation of electromagnetic and gravitational wave observatories will open unprecedented windows to the birth of the first supermassive black holes. This has the potential to reveal their origin and growth in the first billion years, as well as the signatures of their formation history in the local Universe. With this in mind, we outline three key focus areas which will shape research in the next decade and beyond: (1) What were the ``seeds'' of the first quasars; were there multiple channels, and can we differentiate between them, either for high-z objects and/or SMBHs today; how did some reach a billion solar masses before z$\sim7$? (2) How does black hole growth change over cosmic time, and how did the early growth of black holes shape their host galaxies? Conversely, how did the first stars in primordial galaxies influence the conditions for early SMBH growth; what can we learn from intermediate mass black holes (IMBHs) and dwarf galaxies today? (3) Can we unravel the physics of black hole accretion, understanding both inflows and outflows (jets and winds) in the context of the theory of general relativity? Is it valid to use these insights to scale between stellar and supermassive BHs, i.e., is black hole accretion really scale invariant?

In the following, we identify opportunities for the Canadian astronomical community to play a leading role in addressing these issues, in particular by leveraging our strong involvement in the Event Horizon Telescope, the {\it James Webb Space Telescope} ({\it JWST}), {\it Euclid}, the Maunakea Spectroscopic Explorer (MSE), the Thirty Meter Telescope (TMT), the Square Kilometer Array (SKA), the {\it Cosmological Advanced Survey Telescope for Optical and ultraviolet Research} ({\it CASTOR}), and more. We also discuss synergies with future space-based gravitational wave ({\it LISA}) and X-ray (e.g., {\it Athena, Lynx}) observatories, as well as the necessity for collaboration with the stellar and galactic evolution communities to build a complete picture of the birth of SMBHs, and their growth and their influence over the history of the Universe.

\clearpage

%
%

\section{Introduction}

From a wealth of observations,  
we now know conclusively that supermassive black holes (SMBHs), whose masses are millions to billions of times the mass of the Sun, lie at the heart of every massive galaxy. These massive compact objects, so dense that even light cannot escape their gravitational pull, have a profound impact on the formation and structure of their host galaxies, despite being packed into structures smaller than our own Solar System. How does so much mass wind up in such a confined space, and how does this change from the early Universe to today? How do outflows from these comparatively tiny objects seem to control their host galaxy's ability to form stars? What relics of their formation remain in the local Universe, and what can we learn from stellar-mass BHs and extremely high-resolution images of nearby SMBHs? Thanks to the upcoming availability of a host of next-generation multiwavelength and multimessenger facilities, the stage is now set to answer fundamental questions about the nature of SMBHs. In the following white paper, we identify opportunities for the Canadian astronomical community to play a leading role in these discoveries.
\section{How were the Most Massive Black Holes Formed in the Early Universe?}\label{early}

\subsection{Background}

\begin{wrapfigure}{r}{0.6\textwidth}
\begin{center}
\vskip-0.75cm
\includegraphics[width=0.6\columnwidth]{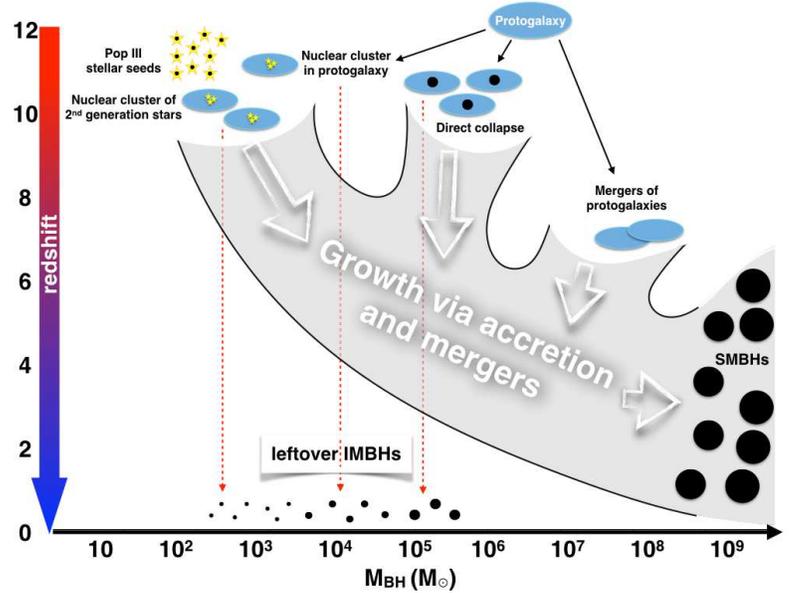}
\caption{Possible channels for the formation of SMBHs, including the epochs in which their ``seeds'' are born, their subsequent evolution with redshift, and the expected masses of ``left-over'' IMBHs to be found in the present era. From \cite{Mezcua2017}.}
\label{fig:Mezcua}
\end{center}
\end{wrapfigure}

While the existence of SMBHs has now been firmly established \citep[see e.g.,][]{FF05}, their origins remain a mystery. How did a black hole in every galaxy become many orders of magnitude more massive than expected for the remnants of even the most massive stars today? A number of possible channels have been proposed to produce the ``seeds'' of massive black holes (BHs) in the early Universe, including primordial formation, the precipitous growth of a single first-generation (Pop III) stellar remnant, or the collapse of a dense stellar cluster or a supermassive star.  
These seeds must then grow via accretion and mergers to become the SMBHs that we observe today \citep{Mezcua2017}. Each of these scenarios (see Fig.~\ref{fig:Mezcua}), as well as some more exotic ones, remain plausible yet unproven \citep{Woods2019}.

The greatest challenge to any theory of SMBH seed formation has been the discovery of quasars with extremely massive BHs ($\sim 10^{9}$--$10^{10}M_{\odot}$) already in place by z~$\sim$~6--7, i.e., well within the first billion years of the Universe \citep[e.g.,][]{Willott2003,Banados2018}. Canada has played a leading role in understanding the connection between the early growth of BHs and galaxies, first by study of luminous $z>6$ quasars from the SDSS \citep{Willott2003}, and then with studies extending to lower luminosity, and BH mass, quasars discovered in the Canada-France-Hawaii Quasar Survey \citep{Willott2007,Willott2010,Willott2013} using CFHT, Gemini, and ALMA.  

How did some BHs become so massive so quickly? The severity of the problem is best illustrated if we consider the time needed for a black hole with initial mass $M_{0}$ to grow to a given mass $M_{\rm{BH}}$ at its (theoretically) maximum rate (i.e., assuming Eddington-limited accretion): $t_{\rm{growth}} \approx 0.45 \frac{\epsilon}{1 - \epsilon}\ln(M_{\rm{BH}}/M_{0}) \rm{Gyr},$
where $M_{0}$ is typically 10--100 $M_{\odot}$ for a Pop III stellar remnant and $\epsilon\sim 0.1$ is the typical radiative efficiency for thin-disk accretion. 
The timescale to grow from a $\sim10M_{\odot}$ BH to 1--10 billion $M_{\odot}$ is then comparable to, or greater than, the age of the Universe at z$\sim$7, pointing to a serious tension between models and observations.


One possible solution is that the radiative efficiency of accretion is much lower for at least some Pop III BHs (so-called ``super-Eddington" accretion). Observationally, there is mixed evidence for the viability of super-Eddington accretion in this case. While narrow-line Seyfert-I AGN are thought to be accreting at high-Eddington rates, \citep[e.g.,][]{2018rnls.confE..34G}, 
AGN are generally found to obey the Eddington limit \citep[e.g.,][]{Peterson04}. Some stellar-mass BHs in X-ray binaries \citep[as well as ULXs, see e.g.,][]{Gladstone2014} are thought to be in the super-Eddington regime \citep[see e.g.,][]{MillerJones2019}, however it remains unclear if this is applicable to high-mass objects (see below). There is a further obstacle presented to stellar-mass (``light'') seeds: while super-Eddington rates appear to be attainable, 
most Pop III BHs are ``born starving'' -- unable to grow rapidly due to strong ionizing radiation, a lack of available gas, and ejection from their hosts via 3-body interactions \citep[see e.g.,][for a review]{Woods2019}

Alternatively, some high-z SMBHs may have formed with much higher initial masses (``heavy'' seeds, with $\sim 10^{4}$--$10^{5}M_{\odot}$) than possible for typical Pop III stellar remnants. Given a number of recent theoretical advances \citep[see e.g.,][for a review]{Woods2019}, primordial atomic-cooling halos have been increasingly favoured as a viable precursor for producing such massive BH seeds. In this scenario, the molecular hydrogen within a primordial halo is dissociated by Lyman-Werner radiation from a nearby stellar population (or populations); this permits the halo gas to reach T$\sim 10^{4}$K, allowing Jeans masses and infall rates of $\sim 10^{5}M_{\odot}$ and $\sim 1M_{\odot}$/yr respectively. The result is the formation of a supermassive protostar, which can collapse due to the post-Newtonian instability to produce a ``direct collapse'' black hole with M$\sim 10^{5}M_{\odot}$ \citep[see, e.g.,][]{Woods2017}. The frequency with which the necessary conditions for such evolution may arise in the early Universe remains poorly constrained, however, 
awaiting both new theoretical insights and the availability of next-generation facilities. 

\subsection{Key Science Goals: 2020-2030 and Beyond}

The next decade will see a revolution in our understanding of the formation and early growth of SMBHs, in particular driven by the availability of transformational new IR observatories. With an anticipated launch in 2022, the {\it Euclid} space telescope will survey most of the high-galactic-latitude sky to discover {\it at least} 100 quasar candidates at 7 $<$ z $<$ 7.5, and 25 beyond the current limit at z$>$7.5. Follow-up spectroscopic confirmation with {\it JWST} as well as existing 8m-class and future 30m-class near-IR telescopes will permit a determination of the quasar luminosity function and active black hole mass function at 7 $<$ z $<$ 8 \citep{Euclid2019}. The relative number densities of high and low mass BHs and the characteristics of their host galaxies with ALMA will constrain the BH growth processes at work (see e.g. Fig.~\ref{fig:Mezcua})

An unambiguous differentiation between ``light'' and ``heavy'' seed models for a given object, however, would require observations at even higher redshifts, as any evidence of the initial formation history of a high-z SMBH would likely be erased by z$\sim$7 following subsequent growth via accretion and mergers \citep[see][for further discussion]{Woods2019}. Emission from a heavy seed accreting at z$\gtrsim$10 is expected to peak at $\approx$1.5$\mu$m, however its total luminosity depends sensitively on a number of factors \citep[time elapsed since formation, metallicity of environment, etc., see][for further discussion]{Pacucci2019}. As an illustrative example, a $10^{5}M_{\odot}$ BH accreting primordial metallicity gas would be robustly detectable by {\it JWST} given a $\sim$ 10ks exposure, if observed within roughly the first $\sim115$Myr of its accretion history \citep{Pacucci2019}. The spatial resolution of {\it JWST} or future 30m-class telescopes are required to determine if the emission originates in a compact region to differentiate it from rest-frame UV stellar emission. Strongly complementary to such efforts would be observations with high-sensitivity next-generation X-ray observatories; depending on the typical column densities, {\it Athena} and/or {\it Lynx} would be able to provide an invaluable check on the emission mechanisms of accreting BH seeds \citep{Pacucci2019}. An alternative strategy may be to seek out the progenitors of light or heavy BH seeds themselves. If heavy seeds are produced via the collapse of truly supermassive stars, they should be detectable (particularly with the aid of gravitational lensing) by {\it JWST} out to z$\lesssim20$, and with {\it Euclid} and {\it WFIRST} out to z$\sim$10--12 \citep[e.g.,][]{Surace2018}. Any such detection would provide clear evidence of the viability of the heavy seed channel for SMBH formation; at the same time, ruling out their existence even within a relatively small volume ($\sim$ several NIRCam fields of view) would begin to rule out the most optimistic heavy seed models \citep[see][]{Habouzit2016}. 
Finally, any formation channel(s) leading to the birth of SMBHs will also produce many seeds that do not reach supermassive size; this should leave behind intermediate-mass BHs (IMBHs, see $\S$\ref{nearby}), whose masses would provide a clear signature today \citep[recall Fig. \ref{fig:Mezcua},][]{Mezcua2017}. 

Canada is uniquely well-placed to play a strong role in unveiling the early history of SMBHs in the next decade. Existing and future ground-based facilities from optical to radio are essential for this work. Our substantial investment in {\it JWST} will provide Guaranteed Time Observations of samples of $z>6$ quasars and their host galaxies, as well as possibilities for General Observer proposals. The collaborative involvement of Canadian scientists in the {\it Euclid} mission will enable Canadians to continue to uncover the most distant accreting SMBHs. Understanding the origin of SMBHs is also deeply connected to several other key questions (e.g., Are there IMBHs? How do SMBHs and galaxies co-evolve?) that will see enormous progress in the next decade, as we outline below.




\section{How have Black Holes Grown from the Early Universe to Today?}\label{evolution}

\subsection{Background}

The majority of SMBH growth occurred between z$\sim$1--3 at a time known as “Cosmic Noon”.  This is clear from a comparison of the observed redshift evolution of quasar luminosities with the masses of the local population of SMBHs \citep[e.g.,][]{YT2002}.  What is not yet clear is what (external or internal) factors  trigger this growth and what impact the growth of SMBHs has on host galaxies. To answer these questions, it is imperative that we obtain robust black hole masses in a large range of galaxies. The most direct method available is to measure their masses via dynamical modelling of the observed kinematics of stars and gas in galactic cores \citep[e.g.,][]{FF05}. This, however, necessitates observations that can resolve the central SMBH's gravitational sphere of influence, $r_{h}=GM_{\rm BH}/\sigma^{2} \sim 11~(M_{\rm BH}/10^{8}M_{\odot})/(\sigma/200 {\, \rm km s^{-1})}^{2}~\rm{pc}$, with $\sigma$ the stellar velocity dispersion. Space-based (e.g., {\it HST}) observations and ground-based 8--10-m facilities equipped with adaptive optics have extended such measurements considerably \citep[up to $\sim$100Mpc for $10^{9}M_{\odot}$ objects, e.g.,][]{2012ApJ...756..179M} but are still constrained to the most massive, local galaxies. 

For more distant AGN, reverberation mapping (RM) provides an ideal tool for measuring the SMBH mass \citep[see, e.g.,][]{Peterson04}.  Under the assumption that any observed changes in the line flux from an AGN's broad line region (BLR) are driven by changes in the disk's continuum flux, the time delay $\Delta t$ between changes in the measured continuum and corresponding variations in any given line flux can provide a distance-independent measure of the size $r_{\rm BLR} = c\Delta t$ of the BLR. Given that the breadth of the selected line is due to Doppler broadening $\Delta v$ of the emission from the surrounding gas bound to the SMBH, the mass of the central SMBH $M_{\rm BH} = f (\Delta v)^2 c\Delta t/G$ can then be estimated. The scaling factor $f$ (of order unity) accounts for the shape, inclination, and kinematics of the BLR. Comparison with other BH mass diagnostics suggests this introduces an intrinsic uncertainty of a factor of $\sim$3 \citep{Onken2004}. The use of RM samples has shown that there is a direct connection between the size of the BLR and the continuum luminosity of AGN allowing for estimates of the BH mass when a multi-epoch observing campaign of sufficient sensitivity for RM is not feasible \citep[see e.g.,][]{VO2009}.  These relations have proven invaluable in measuring the masses of high-z quasars; however, there remain systematic uncertainties in any such method \citep[e.g.,][]{2018NatAs...2...63M}, and the extent to which the nature of the BLRs of AGN may vary with redshift remains uncertain. 

The first robust measurements of local SMBH masses in quiescent galaxies revealed a startling correlation between their masses and the bulk properties of their surrounding stellar populations, leading to the derivation of a number of scaling relations (e.g. $M_{\rm BH}$~--~$L_{\rm sph}$, $M_{\rm BH}$~--~$\sigma$  \citep[see e.g.,][]{FF05}). These scaling relations provided strong evidence that the powerful outflows and high luminosities produced by AGN have a strong influence on the evolution of their host galaxies, intimately linking the growth of SMBHs to the formation of stars and the evolution of galaxies on cosmic 
timescales \citep[see e.g.,][for a review]{HB2014}.
Efforts to trace the impact of this ``quasar feedback" (both radiative and kinetic) on the star formation histories of galaxies at Cosmic Noon in detail \citep[e.g.,][]{Mullaney2015,2019MNRAS.488.4126P}, and to incorporate physically realistic models of AGN feedback in detailed simulations of galaxy evolution (e.g., the FIRE simulations; https://fire.northwestern.edu/) are well underway. Many fundamental questions remain, however, regarding the nature of quasar feedback, its impact on star formation globally and on very local scales, and its role relative to stellar feedback.  Answering these questions will require substantial computational efforts and access to next-generation observatories across the electromagnetic spectrum.

\subsection{Key Science Goals: 2020-2030 and Beyond}

\begin{wrapfigure}{rh}{0.6\textwidth}
\begin{center}
\vskip-1.2cm
\includegraphics[width=0.6\columnwidth]{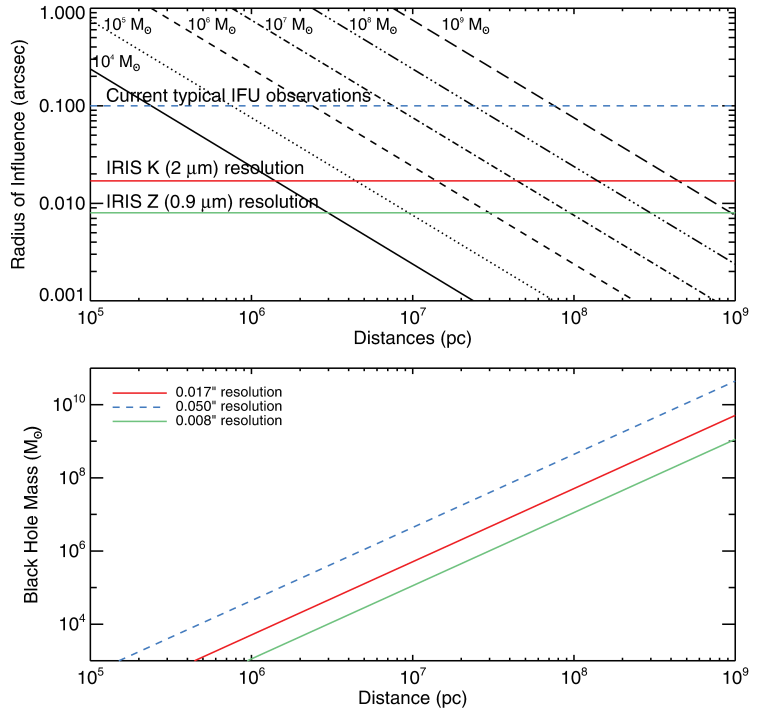}
\caption{Top: Present IFU resolution of the projected radius of influence of SMBHs as a function of distance, contrasted with IRIS capability in K and Z band. 
Diagonal lines denote constant BH mass from $10^{4}$--$10^{9}M_{\odot}$. Bottom: Lower limits on BH mass measurements as a function of distance, for varying angular resolution: typical current IFU (50mas, blue dashed line); TMT's IRIS K band 18 mas (red, solid line); Z band 8mas (green, solid line). From \cite{Do2014}.}\label{fig:IRIS}
\label{fig:TMT}
\end{center}
\end{wrapfigure}

Measuring the growth of SMBHs over cosmic timescales, and disentangling their role in the evolution of galaxies, will require substantial computational efforts and access to ground-breaking new multiwavelength and multimessenger observations. This will be possible in the next decade and shortly thereafter, with the emergence of a host of next-generation facilities. In the near term, the Large Synoptic Survey Telescope (LSST) will, by a variety of selection methods and with the assistance of multiwavelength data, detect $>10^{7}$ quasars (see Fraser et al., LRP 2020 WP). LSST will revolutionize our understanding of the evolution of the AGN luminosity function with redshift (particularly at the faint end of the distribution), as well as studies of AGN variability and clustering, while {\it Athena} and {\it Lynx} will allow us to probe the high-energy emission mechanisms and accretion processes of AGN out to unprecedented distances, building-up a large and complete sample of quasars throughout cosmic history. Large upcoming spectroscopic surveys to be conducted with the Dark Energy Spectroscopic Instrument \citep[DESI; ][]{DESI2016} and Subaru's Prime Focus Spectrograph \citep[PFS; ][]{PFS2014} will facilitate this revolution by providing spectra of millions of quasars at $z>2$.  Other facilities will be essential, however, in order to probe the environments of AGN in detail and provide the high angular and spectroscopic resolution needed for precision measurements of their masses and impact on host environments.

TMT's InfraRed Imaging Spectrograph (IRIS), an adaptive optics fed integral field spectrograph planned for first light, will provide an essential benchmark for such efforts. With IRIS Z-band (0.9$\mu$m) spectroscopy, \cite{Do2014} found that TMT would be able to provide dynamical measurements for over 100,000 SMBHs from the SDSS DR7 sample between 0.005 $< z <$ 0.18, see also Fig. \ref{fig:IRIS}. TMT will also be ground-breaking in studying the host galaxies of SMBHs, capable of measuring the stellar velocity dispersion in galaxies out to z$\sim 1$, and potentially to z$\sim$2-3 using gas kinematics \citep{Ferrarese2010LRP}. Dynamical measurements of SMBH masses in quiescent galaxies will probe the relationship between SMBHs and their hosts in unprecedented detail, while measurements of AGN masses in active galaxies will allow calibration of RM and single-epoch spectral estimates. 

A ground-breaking RM campaign to measure the masses of $\sim$5000 quasars (growing the sample of such measurements by nearly 2 orders of magnitude) would provide an unprecedented look at the innermost parsec of AGNs during the epoch of their peak growth \citep[$z=$1--3;][]{MSE}. As a result, such a legacy survey is a key science goal of MSE. In addition to the tremendous increase in the sample size of reverberation-mapped AGN BH masses, the 11.25-m MSE will offer the high sensitivity required to detect and measure the low-amplitude variability occurring on short timescales in luminous AGN. The high spectrophotometric accuracy of 3-4\% with MSE will enable such low-amplitude variability measurements, subsequently increasing the success rate of the current generation of large-scale AGN RM programs. One of the important MSE specifications that set it apart from other highly multiplexed spectrographs is its wide spectral coverage from 360~nm to 1.8~$\mu$m.  The optical through near-infrared coverage to the H-band will allow simultaneous probing of several lines that span a large range of ionization parameter for reverberation-mapping a wide range of radii during the peak era of SMBH growth. Furthermore, the spectral resolution in the low resolution mode of MSE  is sufficient to obtain velocity-resolved delay maps that can be inverted to trace the structure of the broad-line region for hundreds of AGN.  To date, only a handful of objects have sufficient quality data for such experiments. 

CASTOR, a Canadian-led UV/optical (0.15-0.55~$\mu$m) space mission proposed to the Canadian Space Agency, recently completed a Science Maturation Study (SMS; C\^ot\'e et al. 2019, in preparation) that includes a legacy time-domain survey of thousands of AGN with photometry and grism spectroscopy in two UV bands.  The UV is key for several reasons: (1) The spectral energy distributions of AGN  peak in the UV \citep[e.g.,][]{Shang2011}; (2) UV emission, emerging from the inner regions of the accretion disk, is a more direct probe of the photoionizing continuum close to the central black hole than the optical emission generated further out; and (3) AGN are significantly {\it more variable in the UV} on timescales {\it three times shorter} than in the optical \citep{MacLeod2010}.  One of the key science goals of CASTOR is to study the growth of SMBHs in a wide redshift range (\(0 < z \leq 2.5\)) by measuring BH masses for more than 1000 AGN using the Ly$\alpha\lambda1215$ and CIV$\lambda1548,1550$ lines in the UV.  This would be a significant improvement in the number of BH mass measurements determined from the UV lines that probe the innermost broad-line region which now numbers in the 10s of objects.  The wide redshift space probed by CASTOR will allow direct comparison with and connection to CIV studies at higher redshifts done from the ground (e.g., \(z > 1.5\) with MSE) and Balmer studies at low redshifts (\(z < 1\)).

The next decade will also see remarkable progress in understanding {\it how} the growth of SMBHs over cosmic history is linked with the suppression (or enhancement) of the global star formation rate as a function of time, while also resolving individual distant galaxies in detail to understand how quasar winds and jets propagate through the different phases of their hosts’ ISM. This will be made possible due to the increased sensitivity and resolution of multi-frequency facilities, especially if used in concert \citep[see e.g.,][]{Vayner2017}. 
It is essential for studies of quasar feedback that samples of AGN have not only well-measured dynamical masses, but also host properties (host galaxy mass, SFR etc.) in order to truly trace the effects of SMBH growth and quenching on host galaxies across cosmic time. For this, IFS facilities will be crucial as they will make it possible to map the multi-component gas phases of quasar outflows, as well as contributions from star formation and shocks simultaneously.  JWST’s NIRSpec \& NIRISS will be essential instruments in such an endeavor. The high resolution and $3\times3$ arcsec FOV of NIRSPEC will make possible to trace multiple diagnostics across the entire host galaxy for quasars with redshifts as high as $z\sim2$ (for example, see the JWST ERS proposal Q3D, see http://www.eso.org/~dwylezal/q3d). Additionally, the Canadian NIRISS Unbiased Cluster Survey \citep[CANUCS; ][]{CANUCS2017} will provide resolved emission line maps for thousands galaxies at Cosmic Noon. These results will be particularly meaningful when used in conjunction with high resolution data from ALMA (and eventually ALMA2030), which will simultaneously map the cold molecular gas, the building blocks of star formation. Phase 1 of the Square Kilometer Array (SKA1), meanwhile, will enable a detailed examination of the way star formation and jet power are entwined with the properties of AGN for large samples, and allow for detailed studies of AGN jets up to the epoch of reionization \citep{2015aska.confE..93A}. Further in the future, TMT’s IRIS, in conjunction with Canadian-provided adaptive optics, will make high-resolution studies possible for much larger samples of quasars from the ground.  These observational advances will, in turn, ultimately enable physically-motivated implementations of appropriate resolution for quasar feedback in numerical simulations of galaxy evolution, paving the way toward a deep understanding of the relative importance of AGN feedback in the lives of galaxies.


To understand the growth of SMBHs, we must also understand the physics and demographics of SMBH binaries (SMBHBs) and their mergers. In the coming years, gravitational wave observations will reveal individual SMBHBs, their mergers, and their contribution to the stochastic gravitational wave background. Already, pulsar timing analysis (PTA) has provided meaningful constraints on the masses of any individual nearby SMBHBs \citep{PTAindividual}, as well as the stochastic background \citep{PTAbackground}, and a detection is expected within the next decade.
Canada is playing a vital role in this endeavour through the participation of Canadian institutions in the NANOGrav collaboration using the Arecibo, Green Bank, and JVLA telescopes, and via the CHIME/Pulsar collaboration, which is now carrying out daily observations of all of the millisecond pulsars used in NANOGrav analysis, providing high-cadence monitoring of the intervening ISM (see Stairs et al., LRP 2020 white paper). Going forward, gravitational wave astronomy will play an increasingly prominent role in revealing the origin and growth of SMBHs. With a planned launch in 2034, the Laser Interferometer Space Antenna ({\it LISA}) will be able to detect merging SMBHBs out to $z \sim 15$, tracing the earliest growth of SMBHs, while for more nearby ($z\lesssim 3$) mergers, spin measurements will provide key insight into the relative importance of accretion and mergers in the growth of SMBHs \citep{LISA}. Simply detecting high-z mergers holds the potential to reveal the evolutionary channel(s) leading to SMBH formation \citep[see discussion in][and references therein]{Woods2019}, by revealing the masses of their seeds (recall section \ref{early}), which will also be reflected in the mass distribution of IMBHs (see section \ref{nearby}).




\section{What can we Learn from Black Holes in the ``Nearby'' Universe?}\label{nearby}

\subsection{Background}


Ultimately, the evolutionary mechanisms underlying the formation and growth of SMBHs must be reflected in the observed population in the local Universe. Critically, a still-unknown feature of this population is the distribution, indeed the existence, of intermediate-mass BHs (IMBHs, with M $\sim 10^{2}$--$10^{6}M_{\odot}$). As is clear from $\S$ \ref{early} (recall Fig. \ref{fig:Mezcua}), the mass distribution of any local population of IMBHs (e.g., bimodal, continuous) would be a clear indication of the formation pathways available to SMBHs in the early Universe. Furthermore, understanding the radiative properties of any accreting IMBHs would provide an invaluable benchmark for studies of the epoch of reionisation, BH feedback in the early Universe, and the scale-invariance of BH accretion for arbitrary central mass \citep{Mezcua2017}. Based on hypothesized formation scenarios \citep{Woods2019}, and informed by known mass scaling relations (see $\S$\ref{evolution}), globular clusters (see H{\'e}nault-Brunet et al., LRP 2020, for more) and dwarf galaxies have been a frequent target in the search for IMBHs \citep{Mezcua2017}. Detecting accreting IMBHs in X-ray and radio observations of globular clusters, however, has proven difficult \citep[e.g.,][]{MillerJones2012, Haggard2013}, though this may be understandable given the dearth of gas and dust to be accreted in these environments. Dynamical measurements offer a promising alternative to directly measure the masses of IMBHs, however to date such efforts have largely lacked sufficient resolution to avoid ambiguity in their detections \citep[see][]{Mezcua2017}.

\begin{figure}[th]
    \centering
    \includegraphics[width=0.8\columnwidth]{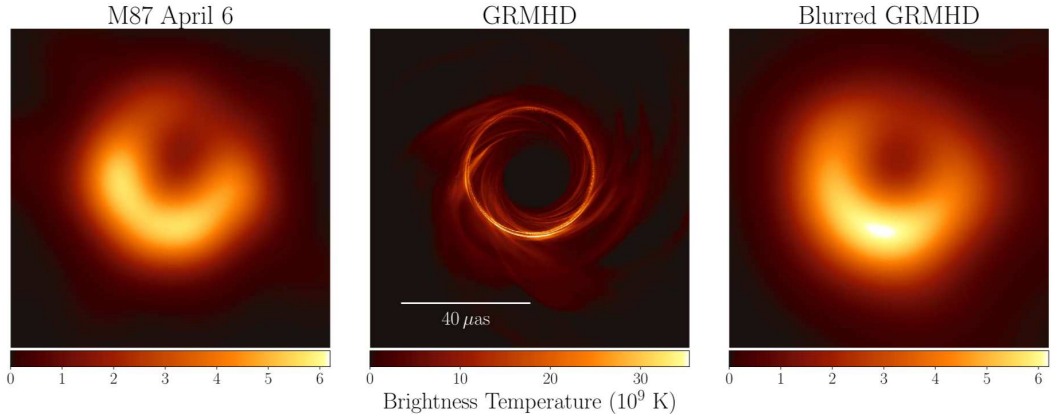}
    \caption{{\it Left}: An {\it Event Horizon Telescope} image of M87's event horizon from Paper I of the discovery series. {\it Middle}: A simulated image based on a GRMHD model. {\it Right}: The model image convolved with a 20 $\mu$as FWHM Gaussian beam. Although the most evident features of the model and data are similar, fine features in the model are not resolved by EHT. From \citet{EHT2019_IV}.}
    \label{fig:eht_m87}
\end{figure}

Increasingly high-resolution IR and sub-mm observations of nearby SMBHs, however, are now beginning to reach all the way down to the event horizon. In late 2018, the GRAVITY experiment \citep{GRAVITY} released the first resolved orbits of hot spots extremely close to Sgr A*'s event horizon, proving for the first time that flares are a horizon-scale phenomenon. In early 2019, the {\it Event Horizon Telescope} provided the first imaging of the last electron and photon orbits near the event horizon of the SMBH M87. These spectacular images 
confirmed our basic prescriptions for how to model MHD turbulence in the presence of a strong gravitational field. 
Canadian scientists made critical contributions to these discoveries, including GRMHD simulations, data modeling and interpretation, and coordinated multi-wavelength observations and acted as essential co-authors on all of the discovery papers \citep[][and associated publications, see also Fig. \ref{fig:eht_m87}]{EHT2019_IV}.
New EHT images illuminating the Milky Way's SMBH, Sgr A*, promise to improve our understanding of Sgr A*'s unusual flares and outflows. These flares occur almost daily at X-ray wavelengths and even more often in the IR, yet the physical processes driving these, and even the radiation mechanisms, are still strongly debated.


Further insight into the nature of the innermost environments of SMBHs has naturally come from high-energy observations. X-ray emission from both the coronae and disks of accreting SMBHs can reveal the local dynamics, the composition of the accreted material, and potentially the mass and spin of the BH. One method for inferring their masses is to invoke the ``Fundamental Plane of BH activity''; a relation between the radio and X-ray luminosities of accreting BHs and the central mass \citep[e.g.,][]{2003MNRAS.343L..59H}. Calibration of this method has relied on measurements of accreting stellar-mass BHs which, although not perfect analogues, allow us to probe accretion and outflows at a variety of Eddington-scaled accretion rates on human timescales. Recent results in accreting, stellar-mass BHs, however, reveal a more complicated picture than when this plane was first discussed, and cross-checking the fundamental plane with other scaling relations, particularly for BH masses thought to be in the IMBH regime, remains an area of active research \citep{2017A&A...601A..20K}. Direct evidence for candidate IMBHs in X-rays has been claimed in the past, from the first observations of ULXs; subsequent investigations have however tended to favour interpreting these objects as super-Eddington accreting stellar-mass BHs \citep{Gladstone2014} or even neutron stars \citep[e.g.,][]{2017A&A...608A..47K}. Going forward, high-energy observations of stellar-mass X-ray binaries will provide an essential probe of extreme accretion rates, essential to understanding the physics underlying the origin, growth, and outflows of SMBHs (recall sections 2 \& 3). Continuing access to the high-energy Universe will be essential in order for Canada to play a leading role in these endeavours. 

\subsection{Key Science Goals: 2020-2030 and Beyond}

Dynamical measurements in the era of 30m telescopes offers a path forward to search for the presence of IMBHs unambiguously, extending the reach of such measurements to $M_{\rm{BH}}\sim 10^{4}M_{\odot}$ at distances of $\sim$ few Mpc (recall Fig. \ref{fig:TMT}). TMT will therefore be essential to push the directly-observed (i.e., via dynamical measurements) SMBH mass function to the IMBH regime in the local Universe \citep{Do2014}, resolving long-standing questions regarding the extension of scaling relations between central BHs and their host galaxy properties to low BH masses in dwarf galaxies, and indeed the formation history of globular clusters \citep{Ferrarese2010LRP, Mezcua2017}.


Canadian astronomers within the EHT collaboration are poised for many breakthroughs using Sgr A* and M87 as laboratories to test fundamental physics. These will arise from multiwavelength coordination with the EHT (submillimeter) and the new GRAVITY experiment (infrared), which will image these SMBHs' accretion flows at scales only a few times their Schwarzschild radii.
Canadians will again play an important role in these investigations via multiwavelength data crucial for interpreting the source variability in the EHT observations. 
With both in-hand and forthcoming observations, we are now on the cusp of creating high-resolution, time-resolved ``movies'' of photons and matter as they approach a BH's horizon, a regime never before probed by observations. Beyond Sgr A* and M87, EHT will soon probe the environment of other extremely massive SMBHs ($>$10$^{10}$ M$_{\odot}$), providing a powerful comparison with Sgr A* (extremely low accretion) and M87 (slightly higher accretion), and new benchmarks for our understanding of accretion. 
Intensive, multi-wavelength, time domain follow-up studies of these nearby SMBHs will be essential to understand the physics of accretion and flares in detail \citep[see, e.g.,][for Sgr A*]{Bouffard2019,Boyce2019,Haggard2019}. The high spatial resolution possible with local BHs can be used to validate interpretation of time domain data that is the only probe of such small scales in more distant objects, allowing calibration of larger samples.   

Ultimately, multiwavelength next-generation facilities will be vital in order to disentangle the environments of nearby SMBHs, as will access to world-class computing facilities in order to interpret and model these results. For example, fostering Canadian leadership in high-energy follow-up of nearby accreting SMBHs, as well as investigations of stellar-mass analogues at all accretion rates, hinges critically on maintaining and growing access to high-energy facilities. Canadians have technically and scientifically contributed to AstroSat, as well as Hitomi and its successor XRISM (to be launched in 2021), and there has been significant growth in high-energy faculty. The Canadian space community is now actively seeking to leverage our existing expertise 
to become a partner in a next-generation X-ray observatory, such as the planned  {\it Athena} mission (see Canadian Space Exploration: Science \& Space Health Priorities 2017). Together with significant multi-wavelength coordination (see Sivakoff, G., LRP 2020 WP), Canada can play a strong role in unraveling the mysteries of BH accretion in the next decade.













\section{Summary \& Priorities}


Understanding the origin, growth, and influence of SMBHs over the life of the Universe is now widely recognized as a fundamental problem in astrophysics, with deep connections to the formation of stars, the evolution of galaxies, and the history of the early Universe. The next decade will see enormous progress in this endeavour, thanks to the availability of revolutionary new multiwavelength and multimessenger facilities. Canada is poised to play an active role in these discoveries thanks to our significant investment in {\it JWST} ($\sim$\$170 million CAD) and TMT ($\sim$\$240 million CAD), as well as our existing optical facilities and our participation in EHT, ALMA, and preparatory developments for the SKA. These facilities will enable Canadian scientists to answer {\bf fundamental questions} regarding the birth of SMBH seeds at high-z, their masses today, and their impact on the gas and dust of their host galaxies over cosmic time. Building a complete picture of these processes, however, will require several additional key investments, planned for the next decade and shortly thereafter. These facilities can provide a clear path for Canada to emerge as a leader not only in SMBH science, but in the astronomical community at large.

From the ground, MSE would provide an unprecedented sample of AGN masses, and would probe their spectral variability on short timescales during the era of their peak growth. This would be a powerful complement to direct dynamical detections of SMBHs from TMT, enabling Canadian astronomers to explore the cosmic evolution of SMBH scaling relations. Given the complementary capabilities in AGN science, and MSE's need for deep, all-sky source catalogs for accurate fibre-placement, there is also a strong synergy with LSST science, accessible to Canadian astronomers with the commitment of a relatively low-cost data processing centre (see Fraser et al., LRP 2020). SKA1 will allow for detailed studies of AGN jets and their impact on their hosts up to the epoch of reionization, and will likely detect gravitational waves from SMBHBs (though prior PTA analysis may do so first, see Stairs et al., LRP 2020 for more details). From space, the next-generation in X-ray observatories will soon transform our view of accreting SMBHs, providing an unprecedented probe of the innermost regions of accretion disks, feedback mechanisms on all scales, and the demographics of SMBHs out to high-z. A Canadian contribution to either {\it Athena} or {\it Lynx} would galvanize high-energy astronomy in Canada in the next decade. The Canadian-led CASTOR mission, meanwhile, would provide an unparalleled tool for echo-mapping of AGN and the evolution of cosmic star formation on sub-galactic scales, and position Canada as {\it the} centre for UV astronomy in the world, for an estimated cost comparable to our 5\% share in {\it JWST}. Notably, space astronomy and instrumentation in Canada is now at a crossroads, for which additional investment is an increasingly existential question.


These facilities present generation-defining opportunities for Canada to play a leading role in answering a number of {\bf priority science questions:} {\bf (1)} What were the ``seeds'' of SMBHs at high-z, and where are their relics (particularly IMBHs) in the local Universe? {\bf (2)} How did the growth of SMBHs change from the early Universe to today, and how did their growth set the course for the evolution of galaxies? {\bf (3)} Can we unravel the physics of black hole accretion, place constraints on the theory of general relativity, and use these to scale between stellar and supermassive BHs? Astonishing progress in understanding these fundamental problems will be made in the next decade, and Canada is uniquely well-placed to take a strong leadership role in this endeavour. 



\begin{lrptextbox}[How does the proposed initiative result in fundamental or transformational advances in our understanding of the Universe?]
\vskip-0.4cm
Understanding how SMBHs are born and the physics of their accretion and outflows is an essential problem, connected to everything from the evolution of galaxies and star formation to probes of fundamental physics. The program outlined above provides a roadmap to address many of these questions in the next decade.

\end{lrptextbox}
\vskip-0.1cm

\begin{lrptextbox}[What are the main scientific risks and how will they be mitigated?]
\vskip-0.4cm
The primary risk is that without access to next-generation facilities across the electromagnetic spectrum, Canada will have to rely on leveraging the resources and expertise it does commit to funding in order to find partners, a precarious position with a likely dimished leadership role.

\end{lrptextbox}
\vskip-0.1cm

\begin{lrptextbox}[Is there the expectation of and capacity for Canadian scientific, technical or strategic leadership?] 
\vskip-0.4cm
There is a strong legacy of Canadian leadership in SMBH science, from uncovering the properties of high-z quasars to revealing the deep connections between SMBHs and their host galaxies. Existing and forthcoming instruments will allow us to lead revolutionary progress in our understanding of SMBHs at all redshifts.

\end{lrptextbox}
\vskip-0.1cm

\begin{lrptextbox}[Is there support from, involvement from, and coordination within the relevant Canadian community and more broadly?] 
\vskip-0.4cm
Canadian support is evident from the broad co-authorship for this WP, with contributors from coast to coast, and from the substantial legacy of Canadian achievements in SMBH science. SMBHs also figure prominently in the science cases for a host of planned international facilities (see Summary \& Priorities, $\S$5).

\end{lrptextbox}
\vskip-0.1cm

\begin{lrptextbox}[Will this program position Canadian astronomy for future opportunities and returns in 2020-2030 or beyond 2030?]
\vskip-0.4cm
There is a strong synergy with future gravitational wave observations (e.g., {\it LISA}, see end of $\S$ \ref{evolution}), and all of the facilities discussed above have broad science cases applicable to a wealth of topics outside of SMBHs.

\end{lrptextbox}
\vskip-0.1cm

\begin{lrptextbox}[In what ways is the cost-benefit ratio, including existing investments and future operating costs, favourable?]
\vskip-0.35cm

Canada is already a substantial partner in many vital upcoming facilities. See also reply to LRP Q.5

\end{lrptextbox}
\vskip-0.1cm

\begin{lrptextbox}[What are the main programmatic risks and how will they be mitigated?
]
\vskip-0.4cm
Construction of TMT and MSE will require an equitable agreement for the future development of Maunakea (see Neilson et al., LRP 2020 WP). Participation in future X-ray facilities remains uncertain, although {\bf some} of the risks could be mitigated by seeking open-access facilities and leveraging Canadian expertise.
\end{lrptextbox}
\vskip-0.1cm

\begin{lrptextbox}[Does the proposed initiative offer specific tangible benefits to Canadians, including but not limited to interdisciplinary research, industry opportunities, HQP training,
EDI,
outreach or education?]
\vskip-0.4cm
Many needs-based opportunities exist for interdisciplinary collaboration with instrumentalists and computer scientists, providing transferable skills for many HQP and the potential for new connections with industry. The planned growth in new facilities must be accompanied by many new faculty hires, providing a strong opportunity to engage in equitable hiring practices, and to stimulate education and outreach in Canada.

\end{lrptextbox}

\end{document}